# Correction of Spherical Aberration of an Immersion Lens Operating Under Space Charge Effect Described by a 2nd Order Equation

**Bilal Khalied Jasem**
*Al-Nahrain University, College of Science, Department of Physics*

**Abstract**

The present work represents a theoretical study for the correction of spherical aberration of an immersion lens of axial symmetry operating under the effect of space charge, represented by a second order function and preassigned magnification conditions in a focusing of high current ion beams. The space charge depends strongly on the value of the ionic beam current which is found to be very effective and represents an important factor effecting the value of spherical aberration .The distribution of the space charge was measured from knowing it's density .It is effect on the trajectory of the ion beam was studied. To obtain the trajectories of the charged particles which satisfy the preassined potential the axial electrostatic potential was represented by a fourth order polynomial function . To measure the optical properties of the lens and to obtain the electrode shape of the electrostatic lens we have solved Poissonequation.

**Keywords**
*Space Charge*
*Spherical Aberration*

# تصحيح الزيغ الكروي لعدسة مغموره تعمل تحت تاثير الشحنة الفراغية الموصوفة بمعادلة من الدرجة الثانية

**بلال خالد جاسم**
*قسم الفيزياء, كلية العلوم, جامعة النهرين/ بغداد-العراق*

**الخلاصة**

يقدم هذا البحث دراسة نظرية في معالجة الزيغ الكروي لعدسة مغمورة متناظرة محوريا تعمل تحت تاثير الشحنه الفراغيه الموصوفة بمعادله من الدرجة الثانية، والتي تعمل تحت ظروف تكبير محددة مسبقا في تبئير الحزم الايونية ذات التيار العالي . ان الشحنة الفراغية تعتمد بشكل كبير على قيمة تيار الحزمة الايونية حيث وجد انها عامل مؤثر ومهم على قيمة الزيغ الكروي. وتم حساب توزيع الشحنة الفراغية من خلال معرفة كثافتها ، وتم دراسة تاثيرها على مسار الحزمة الايونية فوجد انها ذات تاثير مهم على مسارات الحزمه الايونية . وتم تمثيل الجهد الكهروسكوني المحوري باستخدام متعددة حدود من الدرجة الرابعة للحصول على مسار الجسيمات المشحونة الذي يحقق دالة الجهد المقترحة لحساب الخواص البصرية للعدسة من اجل الحصول على شكل الاقطاب للعدسة الكهروسكونية الذي تم خلال حل معادلة بواسون.

## Introduction:

An electrostatic immersion lens consists from two electrodes at a different potential separated by a gap[1] .The focusing action of an immersion lens arises from changes in radial position r and velocity $v_z$ in the gap region. The importance of electrostatic lenses lies in the focusing of accelerated ion beams and are very important part of electron guns and ion source[2]. The theory of charged particles optics of high current beams in which space charge effect are large is important for many applications (microwave oscillator and amplifier tubes) [3].But the space charge effects are more important in heavy ion beams , a heavy charged particles beams are used in such applications like welding ,drilling ,cutting ,etc… [4].

**Space charge effect :**

The space charge effect in a high density ion beams focused by electrostatic lenses is very important . When abeam of charged particles converge to a focus by such lens, collisions between charge particles together with coulomb forces of repulsions tend to diverge the beam and introducing aberrations[5].The spreading of charge particles beam due to its own space charge is connected with collisions between charged particles and negative ions with the wall of the electrodes[3] .The spreading in ion beams is due to the following important effects; i: the effect of negative space charge causes depression of potential in the beam , ii: the repulsive forces between electrons causes spreading of the beam ,and limitation of the beam current[6] .When a large quantity of charged particles are considered , its potential is not negligible compared with that of the electrode [6] .The space charge effect are all based on the potential distribution U(z) inside the beam to determine this distribution the following Poisson's should to be solved [7].

$$\nabla^2 U(r,z) = -\rho(r,z)/\in_0 \text{ ............(1)}$$

The electric potential U(r,z)for a rotationally symmetric electric field with space charge density $\rho(r,z)$ can be expressed as[8].

$$U(r,z) = U(z) - \frac{r^2}{4}\left[U''(z) + \frac{\rho_0(z)}{\in_0}\right] + \frac{r^4}{64}\left[U^{(4)}(z) + \frac{\rho_0''(z)}{\in_0} + \frac{\rho_2(z)}{\in_0}\right] \text{.........(2)}$$

Where U(z) denotes the axial potential along the optical axis and $\in_0$ is the permittivity of a vacuum. and $\rho_0$(z) $\rho_2$(z) are defined by the extended series[8]

$$\rho(r,z) = \rho_0(z) - \frac{r^2}{4}\rho_2(z) + \frac{r^4}{64}\rho_4(z) - \text{......................(3)}$$

The calculation of $\rho(r,z)$ requires the knowledge of many terms of equation (3) .In the present work , for axially symmetric distribution we will consider only $\rho_0(z)$ and $\rho_2(z)$ .The space charge distribution describing the electron beam own space charge for axially symmetric configuration can be expressed by power series[6].

$$\rho(r,z) = \rho_0 + \rho_2 r^2 + \rho_4 r^4 + \text{............(4)}$$

From the equation ,for the current density $J(z)$ [2].

$$J(z) = -\rho(z).v(z) \text{...........................(5)}$$

Where v(z) is the velocity distribution, both distribution depend on the trajectories, the charge distribution $\rho(z)$ is calculated. After some manipulations ,the space charge distribution at a given point of the beam will be determined by the following equation [6].

$$\rho(r,z) = \frac{-I}{S(2\eta U(r,z))^{1/2}} \text{..............(6)}$$

Where I is the total beam current ,S is the area of the uniform cross-section of the beam ,η is the charge to mass quotient of the particles, while the space charge distribution on the optical axis where r = 0 is determined by the following equation [2].

$$\rho(z) = \frac{-I}{S(2\eta U(z))^{1/2}} \text{.......................(7)}$$

For fields having axial symmetry and described by cylindrical coordinates $(r,\theta,z)$,we can write equation (1) in the following from [9].

$$\nabla^2 U(r,z) = \frac{1}{r}\frac{\partial}{\partial r}\left(r\frac{\partial U}{\partial r}\right) + \frac{\partial^2 U}{\partial z^2} = \frac{-\rho(r,z)}{\in_0} \text{..............................(8)}$$

The solution to find $U(r,z)$ in the middle part of equation (8)can be written in a power series[10]

$$U(r,z) = U_0(z) + U_2(z)r^2 + U_4(z)r^4 \text{....(9)}$$

Substitute equation (9) and(4) in equation(8) leads us to find the following new term :

$$\rho 2(r,z) = -\left[\frac{1}{r^2}\rho_0(z) + \in_0\left(\frac{U_0(z)}{r^2} + U_2(z)\right)\right]..$$

.............................................................(10)

Where $\rho_2$(r,z) described the space charge distribution by 2nd order where r is the value. After some trail and error calculations, we have reached the following expression to represent the potential distribution along the optical axis of our immersion lens:

$U(z)=b/2-3/40\ cz + az^2/100 + cz^3/4000 - az^4/20000$ ..........................(11)

Where a, b, and $c$ in eq.(11) are constant .To reduce the spherical aberration effect to its minimum in our immersion lens ,we have to choose values for the applied voltage to the electrodes $V_1$ and $V_2$.To do so, we have done some trail and error calculations. The values that gave minimum spherical aberrations coefficient . $V_1$=4521volt and $V_2$=18926volt.Also we have assumed that Z extends from -10mm to +10mm .The results of the calculations of the potential U(z) as a function of Z using our equation (10) and the field E(z) calculated us the derivative of equation (10) ,are shown in Fig (1).From this fig , we see that the potential is nearly constant at the lens boundaries ,so its first derivatives are zero at the end. This means that there is a field free region out side the lens where the charge particles travel in straight line. Figures(2) and (3) shows respectively the axial space charge distribution $\rho_o(z),\rho_2(z)$ of our immersion lens at various values of beam current .The axial beam distributions at positive values of Z in nearer to optical axis than those at a negative values of Z.

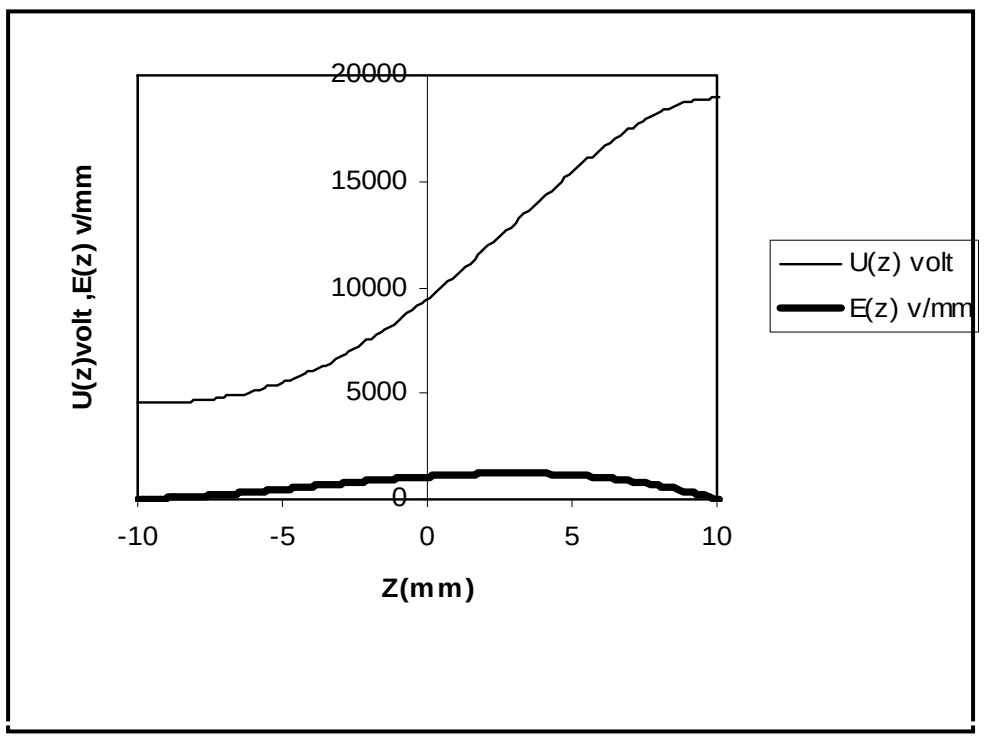

***Fig(1) : The axial potential and field distributions along the optical axis***

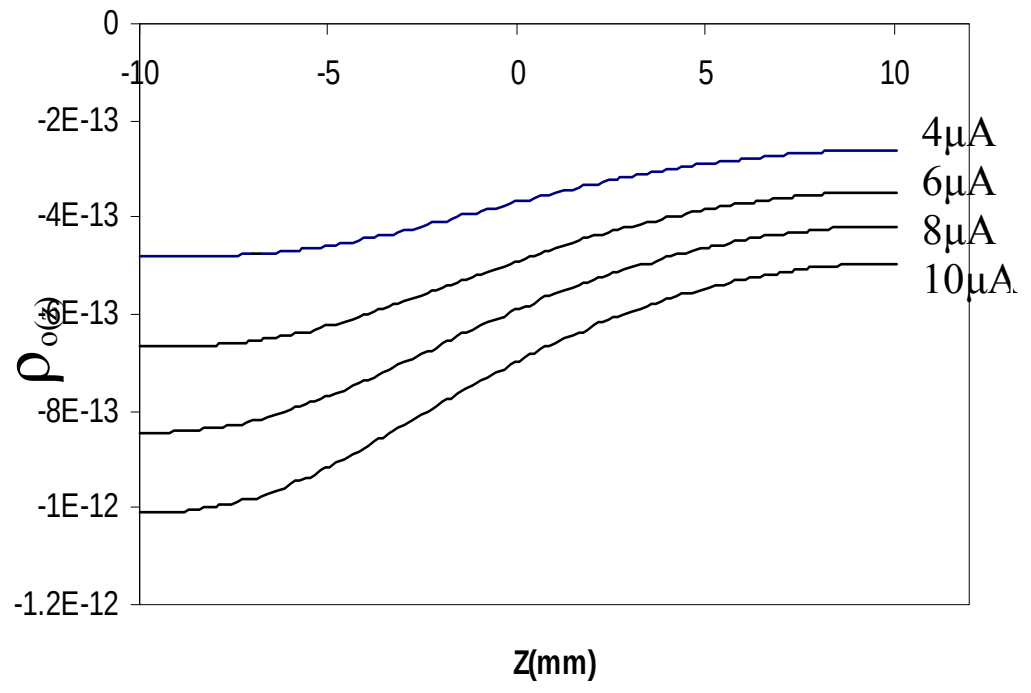

***Fig(2) :The axial charge density $\rho_0$ distribution along the optical axis for different values current***

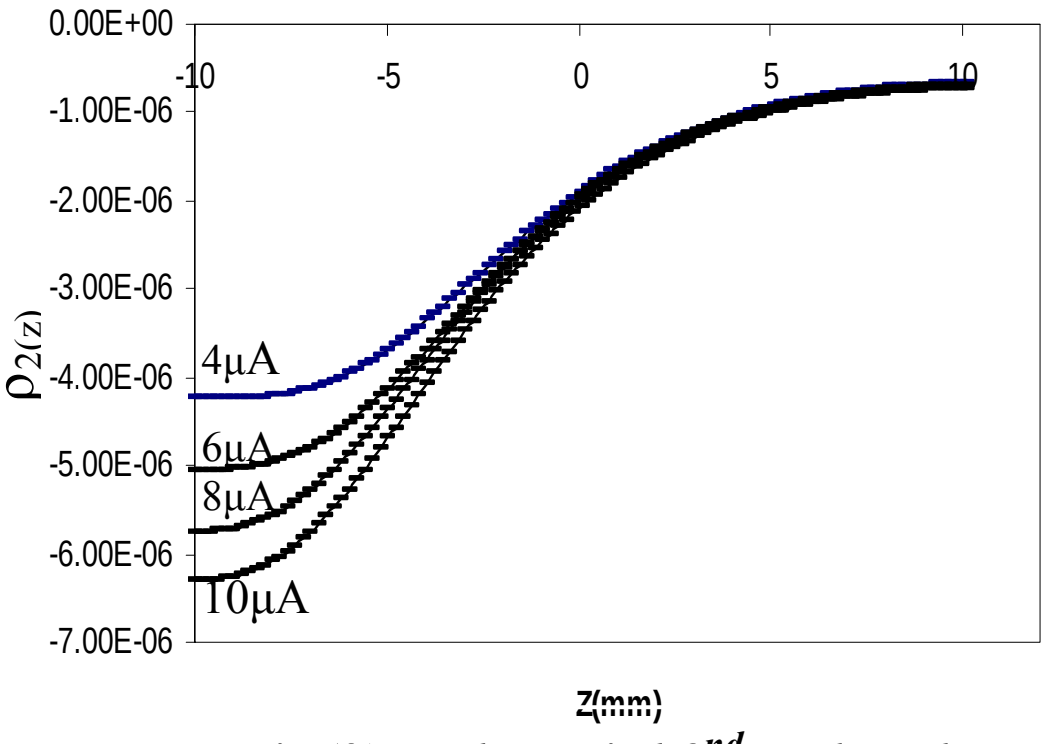

***Fig(3) : The axial 2nd order charge density $\rho_2(r,z)$ distribution along the optical axis for different current values***

The deviation from the optical axis increasing as the beam current increase from 4μA to 10 μA. This is because increasing space charge tends to increase *repulsive power* between charged particles. The *space charge force* defined by $(qr\rho_0/\in_0)$ increase as the radial displacement and beam current. But for the right electrode(positive Z) the distribution is near to the optical axis since the increase in charge density, leads to increase in *the electrostatic radial force* $(qE_r)$.This is because for axial Symmetrical electrode ,this force is inversely proportional to the radial distance $\frac{1}{r}$ [4].

**The trajectory equation in the presence of space charge**

The equation of motion of charged particles traveling at non-relativistic velocity in an axially symmetric electrostatic field in the presence of space charge is given by the fllowing paraxial ray equation [9].

$$r'' + \frac{U'}{2U} r' + \frac{1}{4U}\left[U'' + \frac{\rho_0}{\epsilon_0}\right] r = 0 \quad \ldots\ldots(12)$$

Where U' and U'' are the first and the second derivatives with respect to Z of the axial potential U, respectively. The charged particles beam path along the electrostatic lens field under space charge effect and zero magnification condition is given in Fig (4).this figure shows the trajectories of charged particles beam traversing the electrostatic lens field. Computation have shown that as the beam emerges from the lens field it converges towards but not intersects, the optical axis because space charge effect on the trajectory charged particle also because the term $\left[U'' + \frac{\rho_0}{\epsilon_0}\right]$ in equation (12) is more effective at slope trajectories .

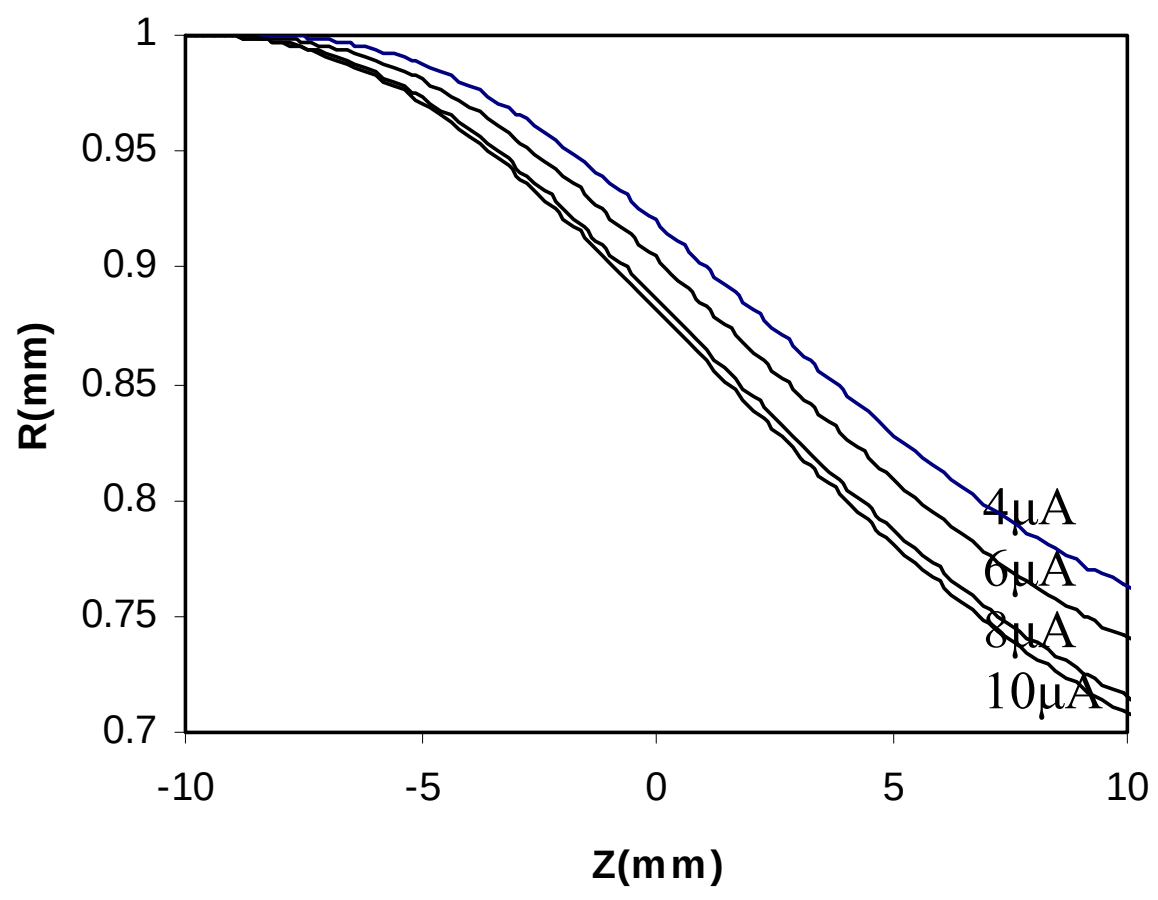

***Fig(4) : Trajectory of charge particles in the electrostatic immersion lens under zero magnification condition at various values of current***

**Spherical Aberration Coefficient**

The most important aberration in an (electron –ion) optical system are spherical and chromatic aberration .The present work focuses on determining only the spherical aberration in the presence of high space charge for an immersion electrostatic lens operated under zero magnification at image side . The spherical aberration coefficient $Cs_i$ at the image plane is calculated from the fallowing equation [13].

$$C_{si} = \frac{1}{16\sqrt{U_i}\, r_i'^4} \int_{z0}^{zi} \left\{ \left[ \frac{5U''^2}{4U^2} - \frac{3U'^2 U''}{8U^3} + \frac{\rho_0}{2\epsilon_0 U^2} \times \left(\frac{\rho_0}{\epsilon_0} + 3U''\right) - \frac{\rho_2 + \rho_0''}{2\epsilon_0 U} \right] r(z)^4 + \left(8\frac{U''}{U} + \frac{3U'U''}{U^2}\right) r'(z) r(z)^3 + \frac{2}{U}\left(\frac{5\rho_0}{\epsilon_0} + 3U''\right) r'^2(z)\, r^2(z) \right\} \sqrt{U}\, dz \quad \ldots\ldots(13)$$

Where U=U(z) is the axial potential ,the primes denote derivative with respect to Z ,and $U_i = U_i(z)$ is the potential at the image side where $z = z_i$ .In the object side ,the spherical aberration coefficient $Cs_o$ is expressed in a similar form to equation (13) where $r_i'^4, \sqrt{U_i}$ are replaced by $r_0'^4, \sqrt{U_0}$ respectively .The integration given in the above equation is executed by means of Simpson 's rule .To correction of spherical aberration in the presence of high space charge is possible by applying high voltage. We have found that the spherical aberration effect in the presence of space charge is very sensitive to variations in the charge density distribution . Fig(5): shows the relative image side spherical aberration coefficient $Csi/f_i$ as a function of Z in the presence of high space charge at various values of beam current for our electrostatic immersion lens operated under zero magnification.

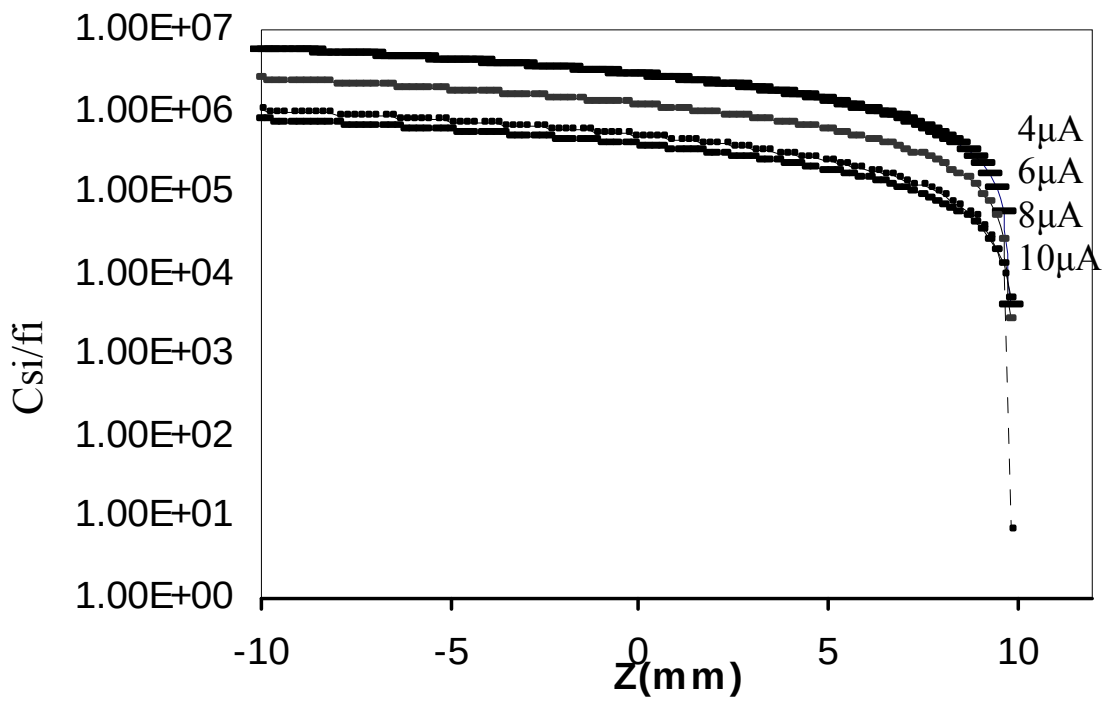

***Fig(5) : The relative image- side spherical aberration coefficients Csi/$f_i$ as a functionof (Z) for an immersion lens at a various values current.***

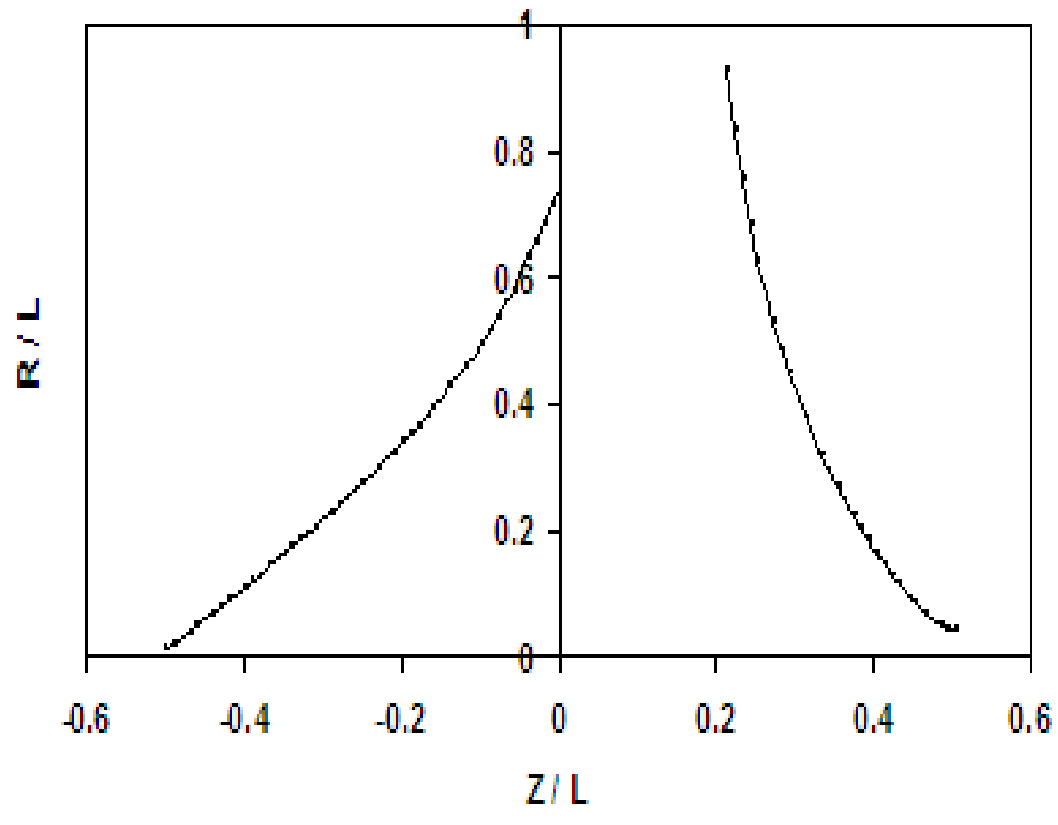

$R_{out}$=0.74L
$R_{in}$=0.017L
Gap=0.23L

$R_{out}$=0.93L
$R_{in}$=0.043L

***Fig(6): the electrodes profile of the electrostatic immersion lens in presence space charge.***

Fig(6): represents the electrodes profile of our electrostatic immersion lens ;the different potential applied to the electrodes depend on the value of a taken into consideration .From the values of the axial potential distribution and its first second derivatives the electrodes profile has been obtained. The radial and the axial dimensions r and z respectively of the electrodes have been normalized in terms of the total lens length which has been taken in the present work to be equal to (20 mm) .

**Conclusion :**

It appears that the proposed analytic equation of the spherical aberration behavior in the presence high space charge for an immersion lens is very sensitive to variation of the charge density distribution along the optical axis .Finding the space charge is highly dependent on the value of the beam current .Which is a very important and effective parameter that affect the aberration of immersion electrostatic lens The effect of space charge on the trajectories of charged particles is clearly apparent from the equation (11) .